\documentstyle[prd,aps,psfig]{revtex}
\include{fleqn}
\include{graphicx}

\def\dlsodos^#1 {{D_{\rm LS}^{#1}\over D_{\rm OS}^{#1}}}
\def\mA{{\cal A}}
\def\mP{{\cal P}}
\def\vn{{\vec n}}
\def\vt{{\vec t}}
\def\ve{{\vec e}}
\def\mg{\big<}
\def\md{\big>}
\def\be{\begin{equation}}
\def\ee{\end{equation}}
\def\ba{\begin{eqnarray}}
\def\ea{\end{eqnarray}}
\def\d{{\rm d}}
\def\xs{x_{\rm str.}}
\def\ys{y_{\rm str.}}
\def\vgamma{\vec{\gamma}}
\def\vdelta{\vec{\delta}}
\def\ncrit{n_{\rm crit.}(x)}
\def\nbcrit{\overline{n}_{\rm crit.}(x)}
\def\Mfunction#1{{\rm #1}}
\def\nb{\overline{n}}
\def\critb{{d_{\rm crit.}}}
\def\lcrit{L_{\rm crit.}}
\def\lcaus{L_{\rm caus.}}
\def\dlcrit{\vec{{\rm d}l}_{\rm crit.}}
\def\dlcaus{\vec{{\rm d}l}_{\rm caus.}}

\def \aa    #1 #2 #3 {{Astr. \& Astrophys. \/} {\bf #1} (#3) {#2}}

\def \aas   #1 #2 #3 {{Astr. Astrophys. Suppl. Ser. \/} {\bf #1} (#3) {#2}}
\def \aj    #1 #2 #3 {{Astron. J. \/} {\bf #1} (#3) {#2}}
\def \apj   #1 #2 #3 {{Astrophys. J. \/} {\bf #1} (#3) {#2}}
\def \apjl  #1 #2 #3 {{Astrophys. J. Lett.\/} {\bf #1} (#3) L{#2}}

\def \apjs  #1 #2 #3 {{Astrophys. J. Suppl. Ser. \/} {\bf #1} (#3) {#2}}
\def \araa  #1 #2 #3 {{Annual Review of Astr. \& Astrophys. \/} {\bf #1} (#3) {#2}}
\def \mnras #1 #2 #3 {{Mon. Not. R. astr. Soc. \/} {\bf #1} (#3) {#2}}

\def \nat   #1 #2 #3 {{Nature \/} {\bf #1} (#3) {#2}}
\def \prevd #1 #2 #3 {{Phys. Rev. D \/} {\bf #1} (#3) {#2}}

\title{Cosmic String Lens Phenomenology: Model of Poisson Energy Distribution}
\author{Francis Bernardeau$^{1}$ and Jean--Philippe Uzan$^{2,3}$}
\address{(1) Service de Physique Th\'eorique, CE de Saclay\\
             F--91191 Gif--sur--Yvette Cedex (France).\\
        (2) Laboratoire de Physique Th\'eorique, UMR--8627 du CNRS,
             universit\'e Paris XI,\\ b\^atiment 210,
             F--91405 Orsay cedex (France)\\
        (3) D\'epartement de Physique Th\'eorique, Universit\'e 
             de Gen\`eve,\\
             24 quai E. Ansermet, CH--1211 Geneva 4 (Switzerland)}
\date{\today}
\draft
\makeatother

\begin{document}

\twocolumn[\hsize\textwidth\columnwidth\hsize\csname @twocolumnfalse\endcsname 
\maketitle
\begin{abstract}     
We present a  novel approach for investigating lens  phenomenology of cosmic
strings  in order  to elaborate  detection strategies  in galaxy  deep field
images. To account  for the complexity of the  projected energy distribution
of string networks we assume their lens  effects to be similar to those of a
straight string carrying a {\em  random} lineic energy distribution. In such
a model we show that, unlike the case of uniform strings, critical phenomena
naturally  appear.  We  explore the  properties  of the  critical lines  and
caustics. In particular, assuming that the energy coherence length along the
string is much smaller than the observation scale, we succeeded in computing
the total length of critical lines per unit string length and found it to be
$4/\sqrt{3}\,{\bf E}(3/4)$.  The length  of the associated caustic lines can
also  be computed to  be $16/(\pi\,\sqrt{3})\,{\bf  E}(3/4)$.  The  picture we
obtain here for  the phenomenology of cosmic string  detection is clearly at
variance with common lore.
\end{abstract}
\pacs{
  {\bf Preprint numbers:} LPT--ORSAY 00/34,
                          UGVA--DPT 00/03-1075, SPhT--Saclay 00/052}
\vspace{0.4cm} ]

\section{Introduction} 

Although the current results on large-scale structure formation favor models
with initial aiabatic scalar perturbation\cite{Inflation}, the formation of
cosmic     strings     is     general     enough    to     merit     further
investigations\cite{Strings}. It is in  particular not excluded that strings
may  form  after  an  inflationary phase\cite{InflationetStrings}.   We  are
interested here in  the phenomenological aspects of cosmic  strings for lens
distortion  effects whereas most  of the  previous investigations  have been
focused on  string detection  from multiple quasar  images\cite{Hogan,Hu} or
Cosmic Microwave Background fluctuations\cite{KS}.   With the advance of new
generation of large  CCD cameras the best direct  evidence for cosmic string
relics is however  likely to be obtained from distortion effects they
induce on background objects such as galaxies.

In a companion paper we insist on the generic properties expected for cosmic
strings  as far  as  lens effects  are  concerned: string  lens effects  are
equivalent to those induced by  a lineic energy distribution. This effective
energy  distribution should  obviously  take into  account  the lens  energy
density, its tension, as well as kinetic energy that might result from rapid
movements or energy currents along the string\cite{peter94b}.

The explicit  computation of  string induced lens  effects has been  done in
various  cases, for  long strings  \cite{Strings}, circular  loops  in plane
transverse  to the  line  of  sight\cite{delaix96}, or  small  loops with  a
multipole  expansion  approach\cite{Hogan}.    However,  much  more  complex
situations   have  been   explored   so  far   with  numerical   experiments
only\cite{delaix97}. In  particular the study  presented in~\cite{delaix97b}
suggests a quite  different lens phenomenology for the  images of background
galaxies but lacks analytical insights. Having in mind such phenomenological
effects,  we  try  in  this  paper  to  get  more  quantitative  results  by
introducing a  simplified description of strings.  In  Sect. \ref{SModel} we
present  the   model  of  ``Poisson''   energy  distribution  we   use.   In
Sect. \ref{ElPhenomenology} general  phenomenological properties observed in
such  a model  are  presented.   Sect.  \ref{StatProp}  is  devoted to  more
precise calculations  on the properties  of the critical lines  and caustics
for such a model of strings and finally we evoque observational aspects.

\section{String model}
\label{SModel}

\subsection{General formalism}

In general lens effects are encoded in an amplification matrix, $\mA$, which
describes the way  the angular positions in the  image plane are transformed
to  those  in the  source  plane  (see  \cite{Mellier} for  a  comprehensive
description  of  lens physics).   It  is usually  written  in  terms of  the
convergence field $\kappa$ and the shear field $(\gamma_1,\gamma_2)$,
\be
\mA=\left(\begin{array}{cc}
1-\kappa-\gamma_1&-\gamma_2\\
-\gamma_2&1-\kappa+\gamma_1
\end{array}\right),
\ee
that are directly  related to the projected energy  density.  The remarkable
property for lens effects are obtained  when one of the eigenvalues of $\mA$
is crossing or getting close to  zero.  The difficulty however is that these
features are neither local nor linear  in the energy density and location of
the  string. The  fact that  the projected  density is  likely  to fluctuate
substantially,  because of local  velocities, wiggles,  longitudinal motions
etc..., has somehow to be incorporated in the description.

In case  of cosmic strings,  the elements of  the deformation matrix  can be
derived the effective projected  potential $\varphi(x,y)$ which can formally
be written  in terms  of the  angular positions $(x,y)$  (e.g. Eq.   (37) of
\cite{BU}] as,
\ba    
\varphi(x,y)&=&4\,G\,\dlsodos^{} \int\d s\ 
\mu[\xs(s),\ys(s)]\times\nonumber\\&&\hspace{-1
cm}\log\left([x-\xs(s)]^2+[y-\ys(s)]^2\right)^{1/2}         
\ea 
where $(\xs(s),\ys(s))$  are the angular string coordinates  for the angular
curvilinear  position  $s$,  $\mu(\xs(s),\ys(s))$  is the  projected  energy
density  at  those  positions,  $G$  is the  Newton  constant  and  ${D_{\rm
LS}/D_{\rm OS}}$ is the ratio of the angular distance between the string and
the source-plane to the one between the observer and the source plane in the
thin  lens\footnote{The thin  lens approximation  is appropriate  if  one is
interested in only a fraction of a  string spanning at most a few degrees on
the sky.  This would  not be appropriate  for apparent string  crossings for
instance.}   approximation\cite{BU}.   The  projected  energy density  is  a
combination of  the projected $T_{00}$, $T_{0z}$ and  $T_{zz}$ components of
the stress-energy tensor of the string if the line-of-sight is along the $z$
direction.  The displacement field is then given by
\be  
\xi_i=-\partial_i\varphi(x,y)  
\ee 
and the elements of the deformation matrix can be written as,
\ba
\gamma_1&=&\left(\partial_x^2-\partial_y^2\right)\ \varphi(x,y),\\
\gamma_2&=&2\,\partial_x\partial_y\ \varphi(x,y),  
\ea 
the local convergence being zero except on the string itself.

\subsection{The Poisson string model}

It seems very  difficult to take into account all the  features that must be
included: the string  are far from being straight  lines with uniform energy
distribution \cite{delaix97}.  We choose to describe  the energy fluctuation
in a  simple manner, assuming that  the string follows a  straight line, but
with local energy fluctuations. This fluctuations are assumed to account for
the  various  changes  of  shape,  density  of  the  strings,  for  possible
non-standard equation of states, or  for the existence of currents along the
string.   We  therefore assume  the  string to  be  straight  along the  $y$
direction,
\be
\xs(s)=0,\ \ \ys(s)=s
\label{straightstring}
\ee
and {\em $\mu(s)$ to be a random  field}. Note that it does not mean however
that the string is actually orthogonal to the line-of-sight. It simply means
that  the string  (or rather  the section  of string  we are  interested in)
appears roughly straight  on the sky. To specify our model  we still need to
explicit  the statistical  properties of  the $\mu$  field.  In  general the
results are going  to depend on the chosen  global statistical properties of
$\mu$ and not only on its  2-point function for instance.  In particular one
question to  ask is whether there  is a finite correlation  length along the
string  or not.  Let  us assume  that the  string  can be  described by  the
following properties,
\ba
\mg\mu(s)\md&=&\mu_0\\
\mg\mu(s_1)\mu(s_2)\md_c&=&c_2\,\mu_0^2\,\exp\left[-{\vert s_1-s_2\vert
\over s_0}\right]\\
\mg\mu(s_1)\dots\mu(s_p)\md_c&=&c_p\,\mu_0^p\,\times\nonumber\\
&&\hspace{-1.5cm}\exp\left[-{{\rm max}\{s_p\}-{\rm min}\{s_p\}\over s_0}\right]
\ea
so that $\mu(s)$ has a finite coherence length, $s_0$, and $\mu(s)$ is
essentially uniform with typical value $\mu_0$ 
at scales smaller than $s_0$. The statistical properties of the displacement
field or of the elements of the amplification matrix are then 
described by the ones of $\mu$. For instance,
\ba
\mg\gamma_i^p(x)\md_c&=&c_p\,G^p\,\dlsodos^p
\mu_0^p\times\nonumber\\
&&\hspace{-2.cm}\int_{-\infty}^{+\infty}\d s_1\dots
\int_{-\infty}^{+\infty}\d s_p\ G_i(x,s_1)\dots
G_i(x,s_p)\times\,\nonumber\\
&&\hspace{-2.cm}
\exp\left[-{{\rm max}\{s_p\}-{\rm min}\{s_p\}\over s_0}\right],
\ea
with 
\ba
G_1(x,y)&=&4\,{y^2-x^2\over \left(y^2+x^2\right)^2},\\
G_2(x,y)&=&-8{xy\over \left(y^2+x^2\right)^2}.
\ea
The general expressions of these integrals are complicated, but
they can be computed at leading order is $s_0/x$. First of all
we can notice that
\be
\mg\gamma_1\md=\mg\gamma_2\md=0.
\ee
At leading order in $s_0/x$ the computation of the
results can be obtained by an expansion of $G_i(x,s_j)$ with
respect to $s_j$ in the vicinity of $s_1$ (for instance) as
\be
G_i(x,s_j)=\sum_p{(s_j-s_1)^p\over p!}\ {\partial^p G_i\over \partial s^p}
(x,s_1).
\ee
It leads to
\ba
\mg\gamma_i^2\md&\approx&8c_2\,G^2\,\dlsodos^2
\mu_0^2\int_{-\infty}^{+\infty}\d s_1
G_i(x,s_1)\times\nonumber\\
&&\hspace{-1.5cm}\int_0^{+\infty}\d s_2\ G_i(x,s_1+s_2)\ \exp(-s_2/s_0)\\
&\approx&8\,c_2\,G^2\,\dlsodos^2
\mu_0^2\,s_0\int_{-\infty}^{+\infty}\d s\,G_i^2(x,s).
\ea
A similar computation can be performed for the third cumulant,
\be
\mg\gamma_i^3\md=
48\,c_3\,G^3\,\dlsodos^3
\mu_0^3\,s_0^2\int_{-\infty}^{+\infty}\d s\,G_i^3(x,s).
\ee
One can further notice that for parity reasons $\mg\gamma_2^3\md=0$.
For the first component of the shear we however have
\ba
\mg\gamma_1^2\md&=&8\,c_2\,s_0\,G^2\,\dlsodos^2 \mu_0^2\,{4\pi\over x^3},\\
\mg\gamma_1^3\md&=&-48\,c_3\,s_0^2\,G^3\,\dlsodos^3
\mu_0^3\,{8\pi\over x^5}.
\ea
One can notice that the dimensionless skewness $\mg\gamma_1^3\md/
\mg\gamma_1^2\md^{3/2}$ is given by
\be
{\mg\gamma_1^3\md\over\mg\gamma_1^2\md^{3/2}}=-
{6\over\sqrt{8\pi}}\left({s_0\over x}\right)^{1/2}{c_3\over c_2^{3/2}}.
\ee
The ratio  $c_3/c_2^{3/2}$ being a  priori finite (and surely  the one-point
probability distribution  function of $\mu$  has no reason to  obey Gaussian
statistics), it implies that the  reduced skewness vanishes at distance much
larger than the correlation length $s_0$  from the string. This is a natural
expectation since at  finite distance and for $s_0\to  0$ such quantities as
the shear  are obtained  as sums  of infinite number  of sources.   So, very
generically, whatever may be the  statistical properties of $\mu$, we expect
that locally,  all quantities  follow Gaussian statistics  as soon  as $x\gg
s_0$.

In the following we will thus assume to be in this regime, that is we assume
that  the energy  distribution along  the string  has a  vanishing coherence
length. In this case the exponential factor can be replaced by a Dirac delta
function so that the 2-point correlation function of $\mu$ can be written,
\be
\mg\mu(s_1)\mu(s_2)\md=2\,c_2\,s_0\,\mu_0^2\,\delta_{\rm Dirac}(s_1-s_2)
\label{mucorr}
\ee
and  the  other  correlation  functions   are  then  not  relevant  for  the
description  of the  distortion  properties.   (It does  not  mean that  the
probability  distribution  function  of  $\mu$  is assumed  to  be  Gaussian
distributed!).  Obviously, this result does not depend on the details of the
small scale correlation properties of $\mu$.  One can see that in this limit
only  one   extra  parameter  is   required  to  describe  $\mu$,   that  is
$c_2\,s_0$. Since  it is  not a priori  possible to distinguish  between the
coherence   length  $s_0$   and  $c_2$,   in  the   following  we   use  Eq.
(\ref{mucorr}) assuming that $c_2=1/2$.

Together with Eq. (\ref{straightstring}) we  call this model of strings, the
{\sl Poisson string  model}. We are aware of  the dramatic simplifications we
have  adopted  to describe  the  outcome of  an  a  priori very  complicated
physics.   Still  we  think  that  some  of the  basic  properties  of  lens
phenomenology can be  captured by such a model in a  much more realistic way
than a simple straight string with a uniform distribution.

\section{Elementary phenomenology of "Poisson string"}
\label{ElPhenomenology}

\begin{figure}
\centering{\psfig{figure=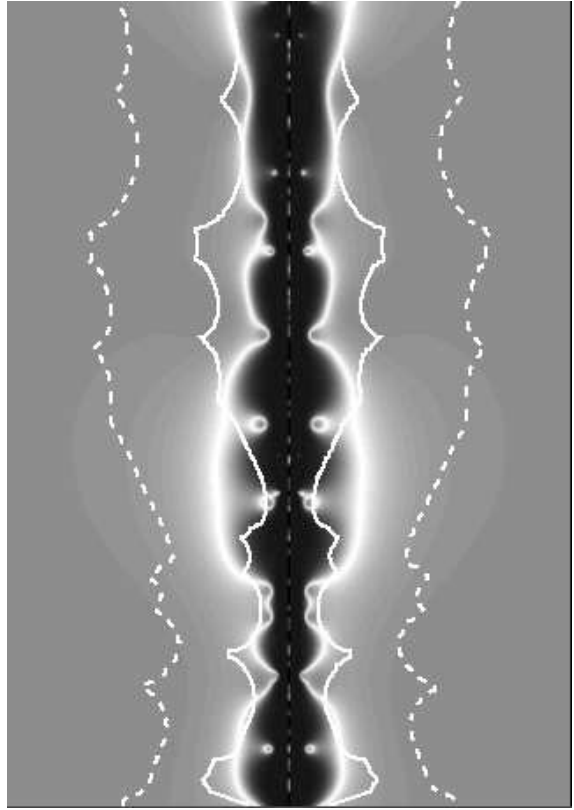,width=7.5cm}}
\vspace{3mm}
\caption
{Numerical experiment showing the  amplification map, i.e. $1/\det(\mA)$, of
a  ``Poisson   string''.   The  brightest  pixels   correspond  to  infinite
magnification: they form the  critical lines.  The darkest pixels correspond
to  a  magnification close  to  zero.  The  solid  lines  correspond to  the
caustics, positions of the critical  lines in the source plane. The external
dashed lines are the counter images of the critical lines.}
\label{AmpMap}
\end{figure}

On Fig.   \ref{AmpMap} we depict  an example of numerical  implementation of
such a  cosmic string,  showing various features  associated with  this lens
system. It  is obviously  beyond the  scope of this  article to  present the
general theory  of gravitational lenses  (see \cite{schneider}). The  aim of
this section is to give some hints on the expected phenomena encountered for
such a model of energy distribution.

 The  grey  levels  show  the   variation  of  amplification  given  by  the
determinant of  $\mA^{-1}$.  Along the brightest areas  the amplification is
infinite; these  locations form the critical  lines. This is  where the most
dramatic lens  effects could be  detected: giant arcs, merging  of images...
Note that such  lines simply do not exist for  strings with uniform density!
The  critical lines  are mainly  made of  two long  lines running  along the
string without  crossing it.  This latter  property is due to  the fact that
whatever  the local  value of  $\mu$ the  displacement field  has  always an
infinite gradient at the string position. This point will be quantified more
precisely in the next section.

Such a  system of  critical lines  is associated with  caustic lines  in the
source plane.  These caustic lines  are shown on Fig.  \ref{AmpMap} as thick
solid  lines:  the  left  one  for  instance is  obtained  by  applying  the
displacement field  to the  right side long  critical line. If  a background
object happens  to lie  on this line,  it will  appear as a  highly deformed
object  (with formally  infinite  magnification) on  the  image plane.   The
caustic  lines  are also  directly  related  to  the possibility  of  having
multiple image of background objects.  Each time one is crossing the caustic
lines the number of associated images is changed by 2: at large distance the
number of image  is one; after one crossing  it has 3 (in other  words if an
object is located in between the  two caustic lines, it has 3 images), after
2 crossings  it has  either 1 again  or 5  (this is  the case  for the
caustic lines associated to the inner critical lines), etc...

Thus, in  general, on  the image  plane the critical  lines are  betrayed by
dramatic lens effects  such as merging of two  images of background objects.
It is then worth noting that we are here generically in a regime of 3 images
in the  vicinity of  the string  (except in rare  cases where  it can  be 5)
instead of  2 as for  a strictly uniform  string. The dashed lines  show the
locations  of the  counter images  of the  critical lines.  It  delimits the
region, in the  image plane, within which multiple images  can be found.  It
can  be  noted however  that  the  amplification  rapidly decreases  in  the
vicinity of the  string, so that central images (i.e.   the ones situated in
between  the  two infinite  critical  lines)  are  expected to  be  strongly
de-amplified, except when they are close to one of the critical lines.

Moreover, the  examination of Fig.  \ref{HSTLens} shows  that numerous small
images generically appear along the string. The presence of these images are
due  to  the fluctuating  small  scale structure  of  the  string. They  are
associated with an infinite number of critical lines (and caustics) near the
string, that are  only partially exhibited on Fig.   \ref{AmpMap} due to the
finite resolution  of our  simulation.  In a  realistic string  these images
would appear down to scales  corresponding to the coherence scale $s_0$. One
aim of the next section is again to quantify more precisely this aspect.

We have checked that, in the limit of the resolution of this simulation, the
global  feature we see  here do  not depend  on the  shape of  the one-point
probability distribution function of $\mu$ neither on the actual resolution.

\section{Statistical properties}
\label{StatProp}

The aim of this section is to quantify some aspects of the results described
in the  previous part.  Indeed, in general systems of  critical lines depend
non-trivially  on  the   lens  strength  and  also  on   the  optical  bench
configuration. So,  at this point, it is  not clear how typical  is the plot
presented  on  Fig. \ref{AmpMap},  and  whether it  depends  or  not on  the
parameter choice. Other aspects that may be difficult to grasp with a simple
numerical  experiment  are  the  mathematical  properties  of  the  critical
lines. Since the  strings has an infinite number  of substructures (when the
coherence length  is assumed to vanish)  the properties of  the critical and
caustic lines may not be numerically  stable. In particular there seem to be
a large  number of small critical  lines close to  the string. So we  do not
know a priori whether those lines  form a fractal like structure, or whether
their total length is finite.

In  the following  we  want in  particular  to locate  the  position of  the
critical lines,  compute their  average length (per  unit string  length) as
well as  the length of their  counterparts in the source  plane, the caustic
lines.  And quite surprisingly, what  in general would be an impossible task
(because critical  lines form complex  patterns that depend  nonlinearly and
nonlocally on the energy distribution) turns out to be computable to a large
extent for this model.

The calculations are  primarily based on the fact  that all local quantities
obey Gaussian statistics and can  thus be entirely described by their second
moments. This property  is valid because we  assume to be in a  regime of an
infinitely small  coherence length on the  string. This would  not have been
the case  otherwise, and  not only should  the correlation effects  of $\mu$
along  the string  be taken  into account  but also  the fact  that $\mu(s)$
itself is not a priori Gaussian distributed.

The calculations of the moments of the displacement, the shear components
and the shear gradients are the required ingredients and are given first.

\subsection{Correlation matrices}

\subsubsection{Displacement field}

Not surprisingly we can check that,
\ba
\mg\xi_1(x)\md&=&-\xi_0\,{\rm sign}(x)\\
\mg\xi_2(x)\md&=&0
\ea
where
\be
\xi_0\equiv4\,\pi\,G\,\dlsodos^{ } \,\mu_0,
\label{xi0def}
\ee
which corresponds to  the case of a uniform cosmic  string. We introduce the
angular  variable $\xi_0$  that  defines  the average  energy  scale of  the
string. The second moment of the displacement field is given by,
\be
\mg\xi_i(x)\xi_j(x)\md={8\pi \over x}s_0\,\mu_0^2\,\delta_{ij}={1\over
2\pi x}s_0\,\xi_0^2\,\delta_{ij}.
\ee
One can note  that the displacement fluctuations have  the same amplitude in
directions  along the string  and perpendicular  to the  string (we  have no
convincing physical interpretation for that). A consequence of these results
is that angular pair separations is  no more fixed, (and given by $2\xi_0$);
they may fluctuate  along the string, and the  amplitude of the fluctuations
is all  the more large that  the images are  close to the string.  The image
pairs  may also  not be  strictly orthogonal  to the  string  direction. The
amplitude  of   these  fluctuations   depend  on  the   dimensionless  ratio
$s_0/\xi_0$. The larger it is, the larger the relative fluctuations are.

Such  effects can be  observed on  Figs. \ref{HSTLens},  where pairs  can be
clearly  identified, but  distances and  orientations indeed  fluctuate from
pair to pair.

\subsubsection{Amplification matrix}

As we already stressed, the main novel effect compared to a uniform straight
string  is the  emergence of  complex distortion  features.  The  latter are
related to  the value  of the local  shear. For  a uniform string  the shear
components are  always zero.  A  reminder of this  property is given  by the
average values of the elements of the amplification matrix that all vanish.

The shear components are explicitly given by
\ba
\gamma_1
&=&8\,G\,\dlsodos^{ } \int\d s\ \mu(s)\,
{(y-s)^2-x^2\over\left((y-s)^2+x^2\right)^2},\\
\gamma_2
&=&-16\,G\,\dlsodos^{ } \int\d s\ \mu(s)\,{(y-s)x\over
\left((y-s)^2+x^2\right)^2}.
\ea
Their second  moments can be calculated\footnote{It is  actually possible to
also   compute   analytically    the   two--point   correlation   functions,
$\mg\gamma_i(x,y)\gamma_j(x',y')\md$,   but   it   has   no   use   in   the
following.}. They depend on the distance to the string and are given by
\be
\mg\gamma_i(x)\gamma_j(x)\md=
{1\over \pi \,x^3}\,s_0\,\xi_0^2\,\delta_{ij}.
\ee
For completeness we can also compute the cross-correlation terms,
\be
\mg\gamma_i(x)\xi_j(x)\md=
-{1\over 4\pi \,x^2}\,s_0\,\xi_0^2\,\delta_{ij},
\ee
with the  displacement field. These results  allow us to  gain some insights
into  this  system.  For instance  one  can  compute  that the  local  shear
distribution function. It is given by,
\be
\mP(\gamma)\d\gamma=\exp\left[-{\gamma^2\over 2\,\sigma_{\gamma}^2}\right]
{\gamma\d\gamma\over\sigma_{\gamma}^2}\label{Pmudmu}
\ee
where 
\be
\sigma^2_{\gamma}=\mg\gamma_1^2\md=\mg\gamma_2^2\md.\label{sgamdef}
\ee
Since the local convergence vanishes, the local amplification
$\mu=1/\det{{\cal A}}$ is given by
\be
\mu={1\over 1-\gamma^2},
\ee
and the critical lines are the locations where $\gamma=1$. From the relation
(\ref{Pmudmu}),  we can  compute the  probability that  a point  at  a given
location is within the critical  zone ($\gamma>1$) or out ($\gamma<1$). This
probability is given by,
\ba
\mP_{\rm crit.}(x)&=&\int_1^{\infty}\mP(\gamma)\d\gamma=\exp\left[-{1\over 2\,\sigma_{\gamma}^2(x)}\right]\nonumber\\
&=&
\exp\left[-{\pi\,x^3\over 2\,s_0\,\xi_0^2}\right].\label{Pcrit}
\ea
This  probability has  a  very simple  shape.  It reaches  unity nearby  the
string. This  expresses the  fact that  the string itself  is always  in the
critical region: no critical lines  can actually cross the string, and there
must always  exist two infinite critical  lines running on each  side of the
string.  The situation  observed  on Fig.  \ref{AmpMap}  is thus  completely
generic in this respect.

This  expression  also  indicates  the  behavior of  typical  critical  line
distance  to the  string. It  must scale  somehow  as $(s_0\,\xi_0^2)^{1/3}$
since this  is the only distance  that intervenes in  Eq. (\ref{Pcrit}). The
exact calculation  of such  a quantity cannot  however be obtained  from the
mere distribution  of the shear.  The number of intersection  points between
the  critical  lines and  any  horizontal lines  for  instance  is also  not
contained in the probability. Somehow  the spatial correlations of the shear
have  to be  taken  into account.  We will  see  in the  following that  the
statistical properties of the shear  gradients allow the computation of such
quantities.

\subsubsection{Shear gradients}

They are a  priori four different shear gradient  components. However, since
the local convergence is zero,  there are simple relations that relate these
components together,
\ba
\partial_x\gamma_2&=&\partial_y\gamma_1,\\
\partial_y\gamma_2&=&-\partial_x\gamma_1,
\ea
so that only two quantities have to be considered. We can furthermore define
a                 pseudo-vector                 $\vdelta$                 as
$\vdelta\equiv(\partial_x\gamma_1,\partial_y\gamma_1)$.  The calculations of
its statistical properties  can be done in a similar  way as previously, and
one gets
\ba
\mg\delta_i(x)\delta_j(x)\md&=&
{3\over \pi \,x^5}\,s_0\,\xi_0^2\,\delta_{ij},\\ 
\mg\gamma_i(x)\delta_j(x)\md&=&
-{3\,\over 2\pi \,x^4}\,s_0\,\xi_0^2\,\delta_{ij}, 
\ea
from which  we deduce  that the correlation  coefficient $r$ with  the shear
field is given by
\ba
r&\equiv& {\mg\gamma_1\delta_1\md\over \sqrt{\mg\delta_1^2\md
\mg\gamma_1^2\md}}={\mg\gamma_2\delta_2\md\over \sqrt{\mg\delta_2^2\md
\mg\gamma_2^2\md}}\nonumber\\
&=&-{\sqrt{3}\over 2}\approx-0.86.\label{rdef}
\ea

As a  result we can explicitly  write the joint  probability distribution of
the local shear and shear gradients,
\ba
&&\mP(\vgamma,\vdelta)\,\d^2\vec\gamma\,\d^2\vec\delta=\nonumber\\
&&\exp\left[-{1\over 2(1-r^2)}\left({\gamma^2\over\sigma_{\gamma}^2}+
{\delta^2\over\sigma_{\delta}^2}-{2\,r\,\vgamma\cdot\vdelta\over
\sigma_{\gamma}\sigma_{\delta}}\right)\right]\nonumber\times\\
&&\ \ \ \ {1\over (1-r^2)(2\pi)^2}{\d^2\vec\gamma\over\sigma_{\gamma}^2}
{\d^2\vec\delta\over\sigma_{\delta}^2},
\label{PDFgd}
\ea
where  $r$  is  given  by  (\ref{rdef}),  $\sigma_{\gamma}$  is  defined  in
Eq.     (\ref{sgamdef})     and     $\sigma_{\delta}$    is     given     by
$\sigma_{\delta}^2=\mg\delta_1^2\md=\mg\delta_2^2\md$.

\subsection{Properties of the critical lines}

\subsubsection{Where are the critical lines?}

\begin{figure}
\centering{\psfig{figure=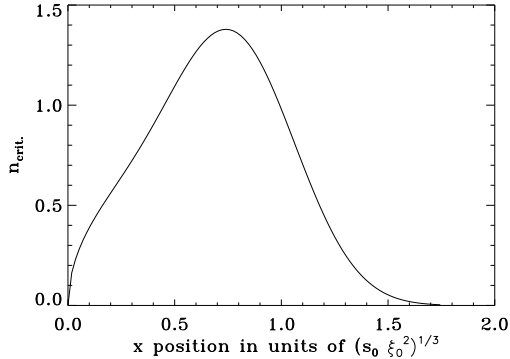,width=7.5cm}}
\caption
{Shape of the function $\nbcrit$ as a function of the distance from
the string in units of $(s_0\xi_0^2)^{1/3}$.}
\label{ncrit}
\end{figure}

As mentionned previously we are now in position to address issues related to
the critical and  caustic lines.  First thing that can  be considered is the
number density,  $\nbcrit$, of intersection point between  the critical line
and any horizontal axis. This function will allow us to compute not only the
typical number  of such  intersection points but  also the  average distance
between the critical lines and  the string, and its fluctuations. The method
we introduce here will also be  employed in the next paragraph to derive the
total length of the critical lines.

If  ${x_i}$   denote  the  locations  where   $\det(\mA)=0$  (which  is
equivalent to  $\gamma=1$ when $\kappa=0$)  on a given horizontal  axis, the
local number density of intersecting points is 
\be 
\ncrit=\sum_i \delta_{\rm Dirac}(x-x_i), 
\ee  
where  each  $x_i$   is  an  implicit  function  of   the  random  variables
$\mu(s)$.  Such a quantity  is actually  a random  quantity that  depends in
particular  on the  $y$ position.  But what  we really  want is  the average
number  density of these  points, that  would be  measured as  a geometrical
average over $y$ position,
\be 
\nbcrit={1\over  L}\int_0^L\d y\,\ncrit, \label{nbcrit}
\ee
in the limit of  a large $L$.  If we assume this  system to be ergodic, then
the geometrical average of Eq. (\ref{nbcrit}) can be replaced by an ensemble
average over  the positions $x_i$  (that are random quantities  depending on
the variables, $\vgamma$, $\vdelta$, etc...) as,
\ba      
\nbcrit&=&\nonumber\\
&&\hspace{-1.5cm}\int\mP(\vgamma,\vdelta,\dots)\d^2\vec\gamma\,
\d^2\vec\delta\,\dots\,\sum_i \delta_{\rm Dirac}(x-x_i).  
\ea
To perform  such an  average, one needs  to make  a change of  variable from
$x_i$ to local parameters $\vgamma$  and $\vdelta$.  This  is possible since
we need  to do this  change of  variable only in  the vicinity of  $x_i$ (we
follow here the same  line of computation as what has been  done in 3D or 2D
Gaussian fields  for computations  of number density  of maxima\cite{bbks}).
The  constraint $(x-x_i)=0$  can then  be replaced  by a  constraint  on the
random  variables  $\vgamma$  and   $\vdelta$.  Since  at  first  order  in
$(x-x_i)$, 
\be \gamma(x)\approx1+(x-x_i)\,\partial_x\gamma(x_i) 
\ee 
we obtain that
\ba
\nbcrit&=&\nonumber\\
&&\hspace{-1.5cm}\int\mP(\vgamma,\vdelta)\d^2\vec\gamma\,\d^2\vec\delta\,
\delta_{\rm Dirac}(\gamma-1)\vert\partial_x\gamma\vert,
\label{expncrit}
\ea
with
\be
\partial_x\gamma={1\over 2}\left(\gamma_1\delta_1+\gamma_2\delta_2\right)=
{1\over 2}\vgamma\cdot\vdelta.
\ee
The integral (\ref{expncrit}) can then be easily computed
explicitly using the expression (\ref{PDFgd}). It leads to
\ba
\nbcrit&=&{{{\sqrt{{\frac{3\,x}{2}}}}}{e^{-2\,\pi\,{x^3}}}} + \\
&&{{{3\over2}\,\pi \,{x^2}\,\Mfunction{Erf}({\sqrt{{\frac{3\,\pi }{2}}}}\,
        {x^{{\frac{3}{2}}}})}{{e^{-{\frac{\pi \,{x^3}}{2}}}}}},
\nonumber
\ea
when  $x$  is  taken  in  units  of  $(s_0\,\xi_0^2)^{1/3}$.   As  shown  on
Fig. \ref{ncrit} where we  have depicted $\nbcrit$, the intersections mostly
take place at a  distance of about $0.75\,(s_0\,\xi_0^2)^{1/3}$. The average
number of intersection points (on one side of string) is then
\be
\nb=\int_0^{+\infty}\d x\,\ncrit={2\over \sqrt{3}}\approx 1.155.
\ee
The number  of intersection points  being an odd  number, it means  that the
critical line crosses  one horizontal line more than once in  at most 7\% of
the  cases. It  supports the  fact that  we are  dominated by  the  two long
critical lines  located on  each side  of the string.  In rare  cases, inner
critical lines can give rise to complex multiple image systems.

The  typical distance  of  the critical  lines  to the  string  can also  be
computed.  It is given by
\be
\critb={\int_0^{+\infty}\d x\,x\,\nbcrit\over
\int_0^{+\infty}\d x\,\nbcrit}
\approx 0.70\,(s_0\,\xi_0^2)^{1/3}
\ee
whereas the scatter of this distance is about
\be
\Delta\critb=0.31\,(s_0\,\xi_0^2)^{1/3}.
\ee
One  recovers   the  scaling  behavior  suggested   in  previous  paragraph.
Remarkably the average  distance and the scatter of  the distance follow the
same  scaling law.  It suggests  that the  critical line  system  is somehow
universal, in the  sense that it does not depend  on the dimensionless ratio
$s_0/\xi_0$. This idea  is further supported in the  next paragraph where we
compute the total length of the critical lines.

\subsubsection{Length of the critical lines}

The length  of the critical lines (per  unit string length $L$)  is a priori
given by
\be
\lcrit=2\int_0^{\infty}\!\d x\int_0^L\!\d y\,\delta_{\rm
Dirac}(x\!-\!x_c,y\!-\!y_c)\label{lcrit1}
\ee
where $(x_c,y_c)$  describes the running  position of the critical  lines in
the $(x,y)$ plane  and the factor 2  accounts to the fact that  we take into
account  both side  of  the string.  To  complete this  calculation one  can
perform a  change of variables from  $(x,y)$ to $(u_c,v_c)$  where $u_c$ and
$v_c$ are the coordinates of a point in a basis given by $(\vn,\vt)$, normal
and tangential vectors  to the critical line at  $(x_c,y_c)$ position.  Then
the integral (\ref{lcrit1}) reads
\be
\lcrit=2\int\d u_c\,\d v_c\,\delta_{\rm Dirac}(u_c).
\ee
As  previously  the $\delta_{\rm  Dirac}$  function  can  be replaced  by  a
constraint on $\gamma$.  In this case the modulus of  the gradient along the
$x$ direction  is replaced  by the  one along the  $\vn$ direction  which is
$\vert\nabla\gamma\vert$  by construction.  Finally  one gets,  returning to
the $x$ and $y$ variables,
\ba
\lcrit&=&2\int_0^{\infty}\d x\int_0^L\d y\,
\int\mP(\vgamma,\vdelta)\,\d^2\vec\gamma\,\d^2\vec\delta\,\times\nonumber\\
&&\vert\nabla\gamma\vert\,\delta_{\rm Dirac}(\gamma-1).
\ea
This can also be computed straightforwardly by noting that
\be
\nabla\gamma={1\over2}\left(\gamma_1\delta_1+\gamma_2\delta_2,
\gamma_1\delta_2-\gamma_2\delta_1\right)
\ee
so that
\be
\vert\nabla\gamma\vert=\gamma\,\delta.
\ee
The integral over  $y$ is then straightforward and gives  a factor $L$.  The
integral  over $\delta$  can be  performed without  too much  difficulty and
gives,
\ba
\lcrit&=&L\,
2{\sqrt{6}\,\pi^2}\times\nonumber\\
&&\hspace{-1.cm}\int_0^{\infty}\d x\,x^{7/2}\,I_0\!\left(-{3\,\pi\,x^3\over
4}\right)\, \exp\!\left(-{5\pi\,x^3\over 4}\right)
\ea
where   $I_0$  is  the   Bessel  function.    Finally,  using   an  integral
representation of $I_0$, one finds
\be
\lcrit={4\over\sqrt{3}}\,{\bf E}\left({3\over4}\right)\,
L\approx 2.80\,L,
\ee
where  ${\bf  E}$  is  the  complete  elliptic  integral.   Remarkably  this
coefficient is independent  of the parameter of the  model. In particular it
does  not  depend  on  the  dimensionless ratio  $s_0/\xi_0$.  Because  this
integral is  finite, it also  proves that the  closed critical lines  have a
finite total length despite the fact that they are in infinite number.

\begin{figure}
\centering{\psfig{figure=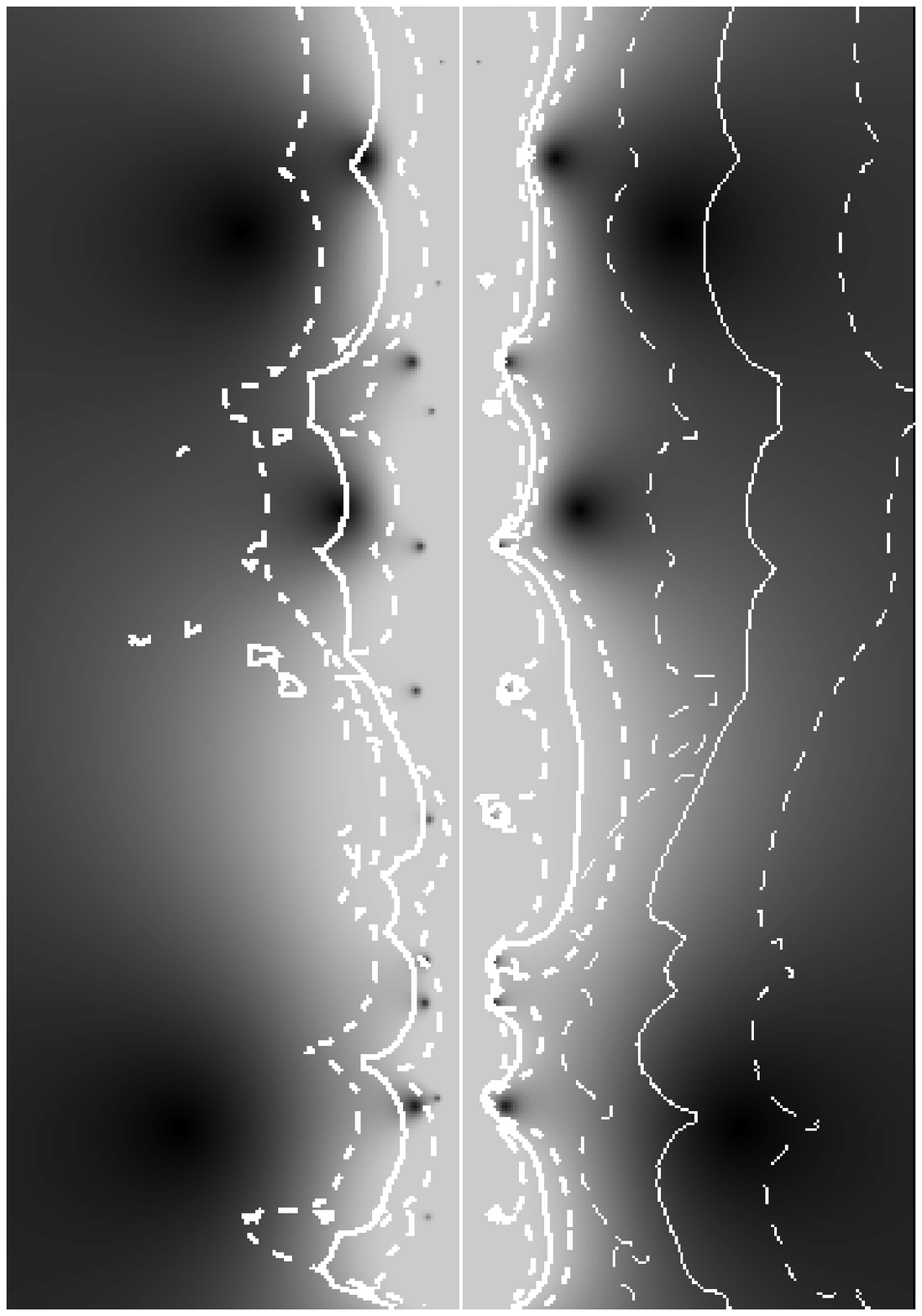,width=7.5cm}}
\vspace{3mm}
\caption
{From the same numerical experiment we show the variation of the location of
the  critical (right side)  and caustic  (left side)  lines when  the source
plane is  displaced. The grey  levels show here  the amplitude of  the shear
field.  We  see that the  critical lines tend  to stick around  points where
$\gamma$ vanishes  (and the points where  it happens are  independent on the
source plane distance). The thin lines show the variation of the position of
the counter image of the caustic lines.  }
\label{CritMap}
\end{figure}

\subsection{Properties of the caustic lines}

If one wants to detect such a string with distortion effects, the efficiency
of such a detection will be  directly proportional to the length of the {\em
caustic}  lines  which are  the  locations on  the  source  plane where  the
background objects are significantly distorted.

The calculation of the  length of the caustic lines is a  priori a much more
difficult task  since it is related  to quantities in the  source plane. One
can  actually circumvent this  problem. The  idea is  that a  length element
$\dlcrit$  on the  critical line  is  going to  be transformed  in a  length
element on the caustic line  $\dlcaus$ according to the amplification matrix
linear transformation
\be
\dlcaus=\mA\cdot\dlcrit.
\ee
We are interested  here only in the change  of length $\vert\dlcrit\vert$ to
$\vert\dlcaus\vert$.   It   depends   locally   on   the   eigenvalues   and
eigen-directions of  $\mA$.  On the  critical line the eigenvalues  of $\mA$
are 0 and 2 (one is necessarily 0 by construction, the other is given by the
trace of the matrix, that is 2, since $\kappa=0$ everywhere). As a result we
deduce that
\be
\vert\dlcaus\vert=2\vert\dlcrit\cdot \ve_2\vert
\ee
where   $\dlcrit\cdot  \ve_2$  is   the  component   of  $\dlcrit$   on  the
eigen-direction   associated  with  the   eigenvalue  2.    The  eigenvector
associated with this eigenvalue is
\be
\ve_2={1\over\sqrt{2(1+\gamma_1)}}\left(\begin{array}{c}
-\gamma_2\\
1+\gamma_1
\end{array}\right).
\ee
If we set
\be
\vgamma=\gamma\,(\cos\psi,\sin\psi)
\ee
and
\be
\nabla\gamma=\gamma\delta\,(\cos\phi,\sin\phi)
\ee
we deduce that
\be
{\vert\dlcaus\vert\over\vert\dlcrit\vert}=
{2\over\sqrt{2(1+\gamma_1)}}\vert\cos(\phi-\psi)+\cos\psi\vert.
\ee
The  shear field  has  no prefered  direction  and thus  it  is possible  to
integrate over $\psi$ with a flat distribution to get
\ba
{\vert\dlcaus\vert\over\vert\dlcrit\vert}&=&\nonumber\\
&&\hspace{-2cm}\int_{-\pi}^{\pi}{\d\psi\over 2\pi}
{\sqrt{2}\over\sqrt{(1+\gamma_1)}}\vert\cos(\phi-\psi)+\cos\psi\vert=
{4\over \pi}
\ea
(this can be  obtained by the change of  variable, $\psi=2\,t$).  Remarkably
this result is independent of $\phi$ so that we directly conclude that
\be
\lcaus={4\over \pi}\,\lcrit\approx3.56\,L.
\ee
>From an observational  point of view we have obtained  with this result some
insights  into the  detectability of  such  a string  from large  distortion
effects.   The  cross-section of  a  Poisson string  of  length  $L$ can  be
estimated to be about $3.6\,L\,$ times typical galaxy sizes, if galaxies are
the objects that one uses to reveal the distortion effects.

\subsection{Cusps number density}

One could  pursue these  investigations by computing  the number  density of
cusps along the critical lines,  that is points corresponding to the merging
of  3  images. These  points  are characterized  by  the  property that  the
critical  line  is  locally   along  the  eigenvector  associated  with  the
eigenvalue  equal to  zero. Technically  it corresponds  to the  cases where
$\vert\dlcaus\vert$  vanishes in previous  calculations. Since  the detailed
calculation  of such  a quantity  requires  one further  derivatives of  the
shear, we do not present it  here.  We can however argue from simple scaling
arguments that
\be
n_{\rm cusps}\propto\,(s_0\,\xi_0^2)^{-1/3}.
\ee

\subsection{Scaling laws}

Note  that for  our description  to be  valid, the  typical distance  of the
critical line to  the string, $(s_0\xi_0^2)^{1/3}$, must be  larger than the
coherence length along the string  $s_0$ (assuming $c_2$ to be finite). This
is the case when the average displacement $\xi_0$ is larger than $s_0$.

Changing  the position of  the source  plane is  equivalent to  changing the
amplitude  of $\xi_0$ while  keeping $s_0$  constant.  It  acts indeed  as a
normalization factor (see Eq. \ref{xi0def}). If this amplitude is multiplied
by $\lambda$, the angular distance of the critical line to the string scales
like,
\be
\critb\propto \lambda^{2/3}.
\ee
But  still the total  length of  the critical  lines remains  unchanged.  On
Fig. 3  we illustrate this effect  by changing the amplitude  of $\xi_0$. We
see that  the critical lines  (thick lines on  the right side)  are drifting
away from  the string. However  they tend to  get stuck on points  where the
local  shear vanishes (dark  patches on  the figure)  creating invaginations
towards the  string position.  When  $\xi_0$ gets large enough  the critical
line is  disconnected and leaves a  small inner closed  critical line around
$\gamma=0$ points. This effect can  be observed in particular in the central
part  of the  figure.  As  a result  the critical  line has  larger position
fluctuations, but  over a larger scale  along the string, so  that its total
length is kept constant. One consequence of this mechanism is that there are
locations  on the  image  plane  (close to  points  where $\gamma=0$)  which
correspond  to quasi-superposition  of  critical lines  from many  different
source planes. That should make them more likely to be detected.

At the same time the caustic lines are sweeping over the source plane; their
distance to the critical line,  and to the string, increases like $\lambda$,
if the typical displacement is larger than $(s_0\xi_0)^{1/3}$.  In this case
we can also observe that the size of the multiple image region is determined
by the displacement  amplitude and thus scales like  $\lambda$ (positions of
the thin lines on the right side).

\section{Discussions}
\label{Discuss}

This  investigation  provides  a  new description  of  the  phenomenological
properties  of lens  physics for  cosmic  strings.  The  model of  ``Poisson
string'' we  present is based on  general results which state  that the lens
effects of string  are those obtained from lineic  energy density.  We think
such a model must have captured  most of the generic properties expected for
string lens phenomenology  as the resemblance with the  results of numerical
experiments (see Fig.  3 of \cite{delaix97b}) strongly suggests  it.  We are
however aware that  the validity of the description we  adopted has not been
demonstrated  (numerical  simulations of  string  networks  anyway lack  the
necessary resolution for describing  correctly the small scale structures of
strings).  It is clear for instance  that the projected position of a string
on  the sky  does not  follow a  straight line.   We have  checked  that the
results we found are not strongly affected if a non-zero curvature radius is
introduced,  as long  as this  radius is  larger than  the  other quantities
involved in this problem, i.e. $\xi_0$ and $(s_0\xi_0^2)^{1/3}$.

\begin{figure*}
\vspace{11.8cm}
\special{vscale=70 hscale=70 voffset=-110 hoffset=-80 psfile=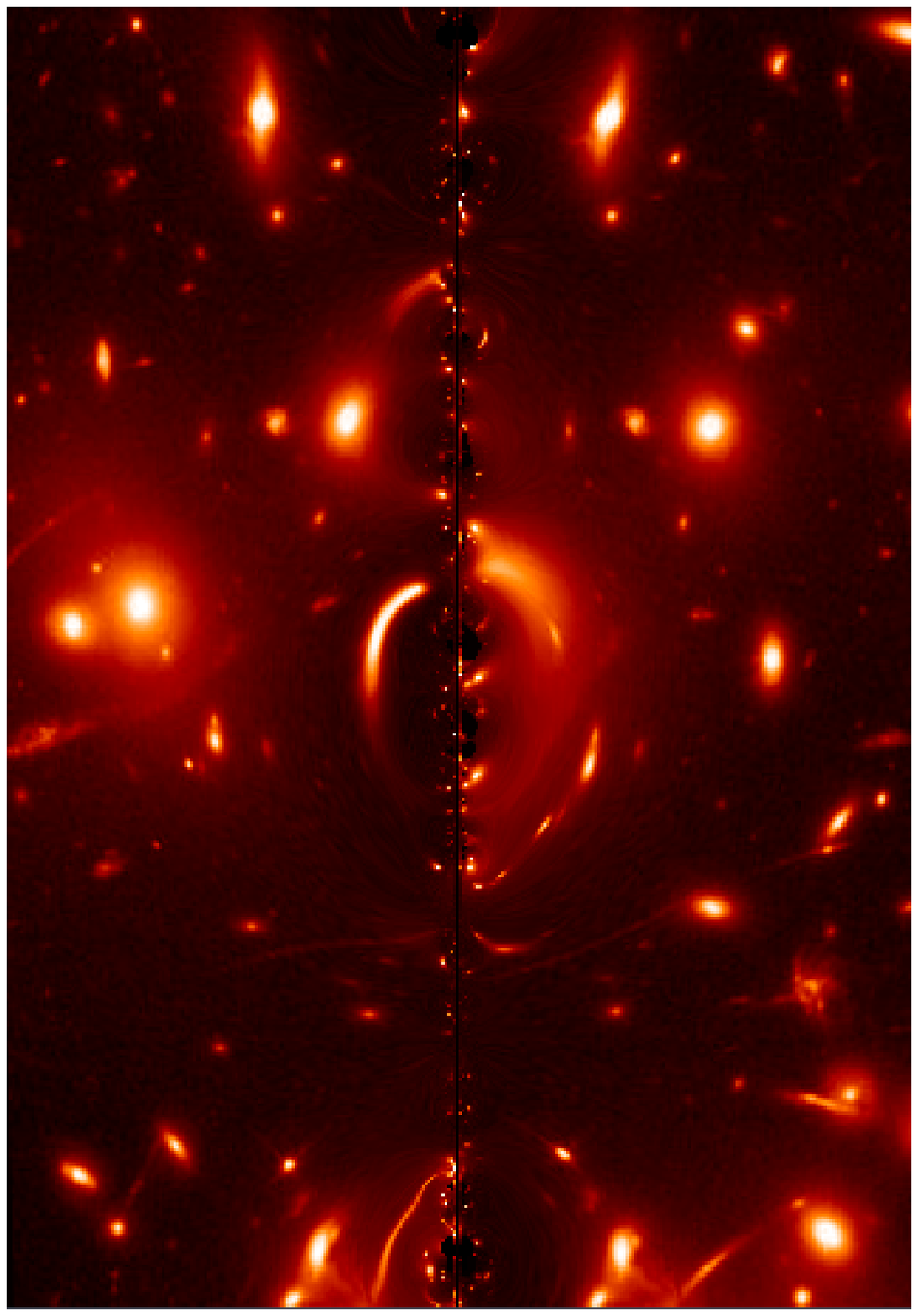}
\special{vscale=70 hscale=70 voffset=-110 hoffset=180 psfile=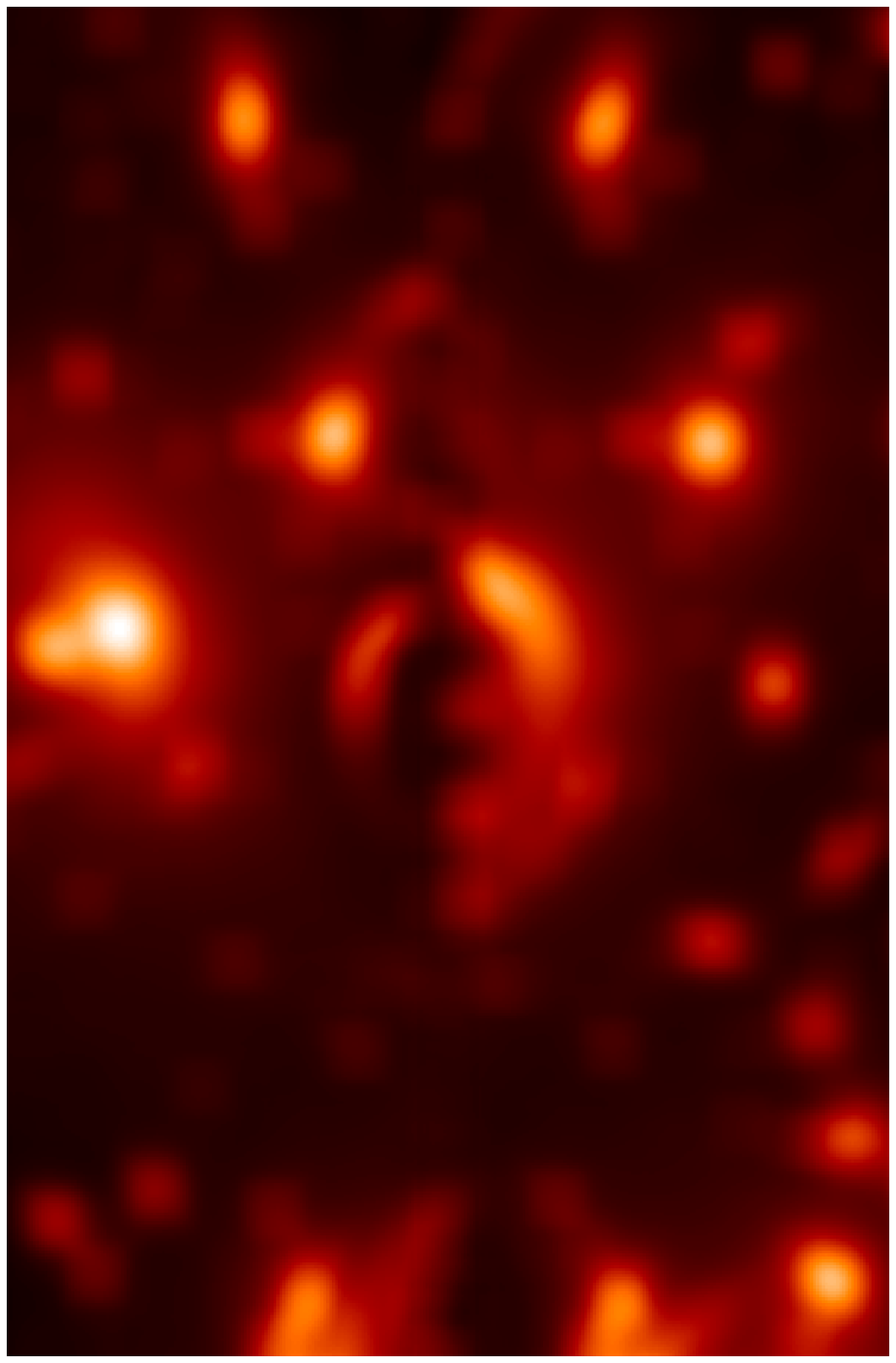}
\caption
{Example of a  deformed image by a Poisson string.  The field corresponds to
an  external   region  of   cluster  A2218  (taken   by  the   Hubble  Space
Telescope). If such a field was put at $z=1$, then an intercepting string at
$z\sim0.8$ along the line-of-sight would produce multiple images as observed
on the 2 pictures (the typical pair separation is about 5 arcsec). The right
panel shows this  system with a resolution of about  0.5 arcsec, whereas the
left panel shows the same image with sub 0.1 arcsec resolution. In this case
the distortions of the background  galaxies are clearly exhibited. Note also
that a  number of small images  appear along the string.  In high resolution
images, that might be the most effective way of detecting strings.}
\label{HSTLens}
\end{figure*}

The results we have obtained suggest novel strategies for detecting strings:
multiple images are still present, but images can have undergone significant
distortions;  small images  of  background  objects can  be  found in  large
numbers...  The most  adapted strategy depends however on  the resolution of
the  imaging  apparatus  and  on   the  energy  scale  of  the  string.   On
Fig. \ref{HSTLens}  we show an  example of an  HST deformed image.   In this
case the energy scale of the string expressed in units of $\xi_0$ is similar
to the angular scale of a galaxy.   For GUT strings, $\xi_0$ would be of the
order  of a  few  arcsecs which  is indeed  a  typical galaxy  size in  deep
surveys. The probability of observing such an event depends obviously on the
survey coverage. The expected angular length of strings to be present within
redshift unity  has been estimated in  \cite{hindmarsh} and is  such that on
average one  string is  expected to  cross a 100  deg$^2$ survey.   But even
though such  surveys are  within reach in  the near future,  the probability
with which one can detect a  string (or conversely the rejection one can put
on  the existence  of  strings with  a  given energy  scale) still  deserves
further investigations in  particular on the effects of  multi source planes
(the amplitude  of the lens  effects depends on  the position of  the source
plane) that may partly blur the searched effects.

\twocolumn[\hsize\textwidth\columnwidth\hsize\csname @twocolumnfalse\endcsname 
]

\end{document}